\documentclass[
aip,groupedaddress,
 amsmath,amssymb,
 reprint,%
]{revtex4-1}

\usepackage{graphicx}
\usepackage{dcolumn}
\usepackage{bm}

\usepackage[utf8]{inputenc}
\usepackage[T1]{fontenc}
\usepackage{mathptmx}
\usepackage{etoolbox}

\usepackage{hyperref}

\makeatletter
\def\@email#1#2{%
 \endgroup
 \patchcmd{\titleblock@produce}
  {\frontmatter@RRAPformat}
  {\frontmatter@RRAPformat{\produce@RRAP{*#1\href{mailto:#2}{#2}}}\frontmatter@RRAPformat}
  {}{}
}
\makeatother
\begin{document}

\title[]{Combining graph deep learning and London dispersion interatomic potentials: A case study on pnictogen chalcohalides \\ \hfill}
\author{ \c{C}etin K{\i}l{\i}\c{c} }
\author{S\"{u}meyra G\"{u}ler-K{\i}l{\i}\c{c}}
\affiliation{
Department of Physics, Gebze Institute of Technology, Gebze, Kocaeli 41400, T\"{u}rkiye
}
\email[ E-mail: ]{cetin\_kilic@gtu.edu.tr}


\begin{abstract}
\centerline{\sl Published version available at \url{https://doi.org/10.1063/5.0237101}}
\vspace*{6pt}
Machine-learning interatomic potential models based on graph neural network architectures have the potential to make atomistic materials modeling widely accessible due to their computational efficiency, scalability, and broad applicability. The training datasets for many such models are derived from density-functional theory calculations, typically using a semilocal exchange-correlation functional. As a result, long-range interactions such as London dispersion are often missing in these models. We investigate whether this missing component can be addressed by combining a graph deep learning potential with semiempirical dispersion models. We assess this combination by deriving the equations of state for layered pnictogen chalcohalides BiTeBr and BiTeI and performing crystal structure optimizations for a broader set of V-VI-VII compounds with various stoichiometries, many of which possess van der Waals gaps. We characterize the optimized crystal structures by calculating their X-ray diffraction patterns and radial distribution function histograms, which are also used to compute Earth mover's distances to quantify the dissimilarity between the optimized and corresponding experimental structures. We find that dispersion-corrected graph deep learning potentials generally (though not universally) provide a more realistic description of these compounds due to the inclusion of van der Waals attractions. In particular, their use results in systematic improvements in predicting not only the van der Waals gap but also the layer thickness in layered V-VI-VII compounds. Our results demonstrate that the combined potentials studied here, derived from a straightforward approach that neither requires fine-tuning the training nor refitting the potential parameters, can significantly improve the description of layered polar crystals.
\end{abstract}

\maketitle


\section{\label{sec:level1}INTRODUCTION\protect\\ }

In atomistic simulation of materials, 
  machine-learning interatomic potentials (MLIPs) offer a cost-effective approach 
  akin to traditional empirical analytical potentials, 
  yet they can deliver first-principles-level accuracy
  enabling realistic simulations over extensive time and length scales. 
  \cite{unke2021machine}
However, 
  like empirical potentials, 
  the applicability of MLIPs can be hindered by transferability issues, 
  particularly when training is restricted to a narrow chemical space with limited data selection.
Addressing this challenge has recently become a focal point of research, 
  encouraging multiple research groups to develop universal MLIP models using graph neural network architectures,
  such as 
      M3GNet (materials 3-body graph network),
      \cite{chen2022universal}
      PFP (preferred potential),
      \cite{takamoto2022towards}
      ALIGNN-FF (atomistic line graph neural network-based force field),
      \cite{choudhary2023unified}
      CHGNet (crystal hamiltonian graph neural network), 
      \cite{deng2023chgnet}
      GNoME (graph networks for materials exploration),
      \cite{merchant2023scaling}
      MACE-MP-0, 
      \cite{batatia2023foundation} and
      MatterSim.
      \cite{yang2024mattersim}
These models are trained on vast datasets of atomic configurations,
  generated through crystal structure optimizations performed via density-functional theory (DFT) calculations,
  involving a multitude of chemical elements from 
  many rows and columns of the periodic table.

The development of MLIP models with general applicability comparable to DFT-based methods
  has the potential to make atomistic materials modeling widely accessible,
  thanks to the computational efficiency and scalability of MLIPs.
Achieving this level of generality and 
                        transferability across numerous chemical elements in combination
  is quite unlikely with any class of analytical potentials.
  \cite{becker2013considerations}
On the other hand, the training datasets for many MLIP models are typically derived from DFT calculations 
  using a semilocal exchange-correlation functional, i.e., 
  the functional of Perdew, Burke, and Ernzerhof (PBE). \cite{perdew1996generalized}
Consequently, these models largely lack long-range interactions.
To address this, some models (PFP and MACE-MP-0) are supplemented 
  with an additive dispersion correction (DFT-D3) potential \cite{grimme2010consistent,grimme2011effect}
  to account for London dispersion interactions.
Despite being weaker than ionic or covalent bonds, 
  dispersion interactions provide attractive van der Waals forces between the layers in layered crystals, 
  cumulatively determining the stability of the crystal.
The inclusion of London dispersion is essential 
  not only for crystal structure optimization 
  but also for the prediction of anisotropy, compressibility, and indirectly, electronic structure, 
  particularly for layered polar compounds such as BiTeI. \cite{guler2015crystal}
Furthermore, London dispersion is critical for developing a universal, broadly applicable potential, 
  even for non-layered crystals, 
  as long as the possibility of a phase transition to a layered phase cannot be excluded \textit{a priori}.

In this paper, we combine the M3GNet potential \cite{chen2022universal} (trained to reproduce PBE/PBE+$U$ energies, forces, and stresses) 
  with the DFT-D3 \cite{grimme2010consistent,grimme2011effect} and DFT-D4 \cite{caldeweyher2020extension} models
  into dispersion-corrected potentials, termed M3GNet+D3 and M3GNet+D4, respectively.
We explore whether these combined potentials provide an adequate physical description of pnictogen chalcohalides
  M$^{\text{V}}$Q$^{\text{VI}}$X$^{\text{VII}}$, where M = As, Sb, Bi; Q = O, S, Se, Te; X = F, Cl, Br, I.
First, we focus on predicting the equation of state (EOS) of BiTeI and BiTeBr,
  which are poorly described by PBE 
    but are well   described \cite{guler2016pressure,sans2016structural} using the PBEsol functional. \cite{perdew2008restoring}
We therefore investigate whether a MLIP trained to reproduce PBE results can be enhanced with dispersion corrections 
  to achieve accuracy comparable to that of PBEsol.
Second, we perform crystal structure optimizations,
  using the M3GNet, M3GNet+D3, and M3GNet+D4 potentials,
  for a set of V-VI-VII compounds with various stoichiometries,
  for which experimental crystal structures are available in the Crystallography Open Database (COD). \cite{gravzulis2009crystallography}
Following these optimizations, we calculate the X-ray diffraction (XRD) patterns and 
  radial distribution function (RDF) histograms
  for the optimized and corresponding experimental crystal structures.
The Earth mover's distance (EMD) \cite{iwasaki2017comparison} between the optimized and experimental structures
  is then computed from the XRD patterns and separately from the RDF histograms.
This approach quantifies the dissimilarity of the optimized structure relative to the experimental structure using two distinct EMDs: 
  one capturing differences in long-range periodicity and interplanar distances and 
  another that is more sensitive to local atomic environments and bond lengths.

Our interest in V-VI-VII compounds is two-fold.
First, they are promising candidates for energy applications due to their appealing electronic and optical properties, 
  along with their wide compositional variety and tunability. \cite{wlazlak2018heavy,ghorpade2022emerging,choi2023heavy}
Enabling cost-effective and realistic atomistic simulation of these compounds and their derived systems 
  via M3GNet+D3 or M3GNet+D4
  would facilitate theoretical studies for understanding their fundamental properties and 
  optimizing their performance in applications.
Second, many of these compounds possess van der Waals (vdW) gaps, either between planes or chains in their crystal structures.
The effect of pressure on the atomic structure of layered V-VI-VII compounds depends strongly on the reduction of the vdW gap. 
Predicting the lattice parameters of layered crystals with vdW gaps
  (and their variation under pressure)
  using the PBE functional, \cite{perdew1996generalized}
  or even a nonempirical van der Waals functional like optB86b-vdW, \cite{klimevs2011van}
  can yield inaccurate,
  or even unrealistic, results. \cite{guler2016pressure}
Accordingly, crystal structure optimizations for pnictogen chalcohalides
  offer a sensible case study for assessing the impact of adding London dispersion corrections to a MLIP.

Our results indicate that the dispersion-corrected M3GNet potentials
  generally (although \textit{not universally}) provide 
  a substantially improved description of the crystal structures of pnictogen chalcohalides.
The improvement, however, becomes less effective for compressed structures, 
  likely because the M3GNet potential was primarily trained on low-energy configurations, 
  which could hinder the broader applicability of the dispersion-corrected M3GNet potentials.
Nonetheless, while M3GNet substantially overestimates the unit cell volume, 
  this error becomes a slight underestimation when D3 and D4 corrections are included.
Importantly, incorporating London dispersion corrections into the M3GNet potential 
  leads to systematic improvements in predicting not only the van der Waals gap but also layer thickness in layered V-VI-VII compounds.
Notably, these significant improvements are achieved with a simple combination 
  that does not require fine-tuning the training process or refitting any potential parameters.

The rest of the paper is organized as follows: 
Section~\ref{s:met} describes the method of calculation and summarizes the computational details.
Section~\ref{s:res} discusses the calculation results, and concluding remarks are presented in Sec.~\ref{s:con}.

%
%
\section{\label{s:met}Method}

\begin{figure*}
  \includegraphics[width=0.78\textwidth]{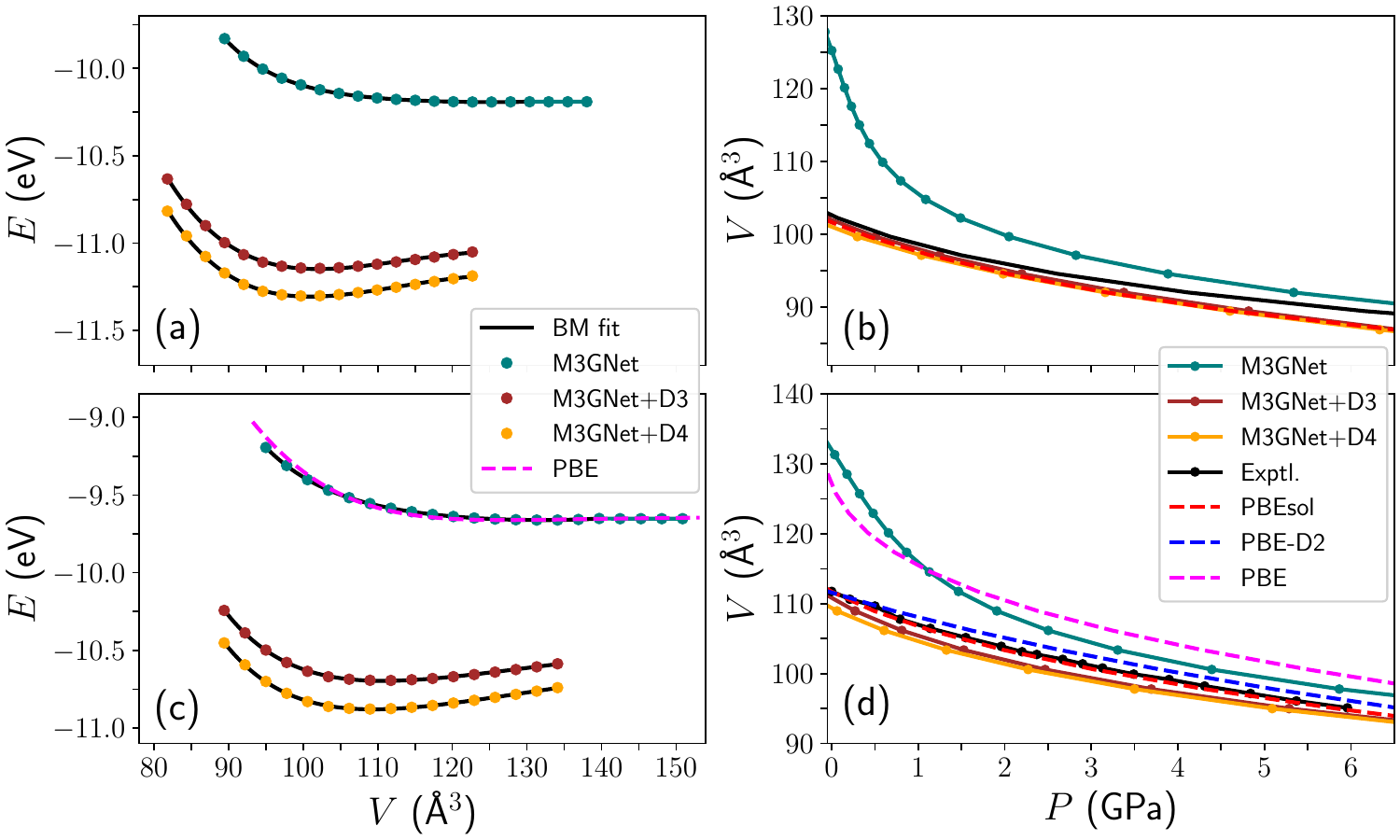}
  \caption{The plots of energy $E$ versus the volume $V$ for BiTeBr (a) and BiTeI (c).
           The black solid curves represent fourth-order Birch-Murnaghan (BM) fits.
           The corresponding volume versus pressure $P$ curves are shown in (b) and (d).
           For BiTeBr, the experimental and PBEsol curves are reproduced using third-order BM fit parameters 
             reported in Ref.~\onlinecite{sans2016structural}.
           For BiTeI, the PBEsol, PBE-D2, and PBE curves are taken from Refs.~\onlinecite{guler2015crystal,guler2016pressure},
             while the experimental curve represents data from Ref.~\onlinecite{xi2013signatures}.}
    \label{f:ev}
\end{figure*}

We combined the M3GNet potential \cite{chen2022universal}
  with the DFT-D3 \cite{grimme2010consistent,grimme2011effect} and 
           DFT-D4 \cite{caldeweyher2020extension} models
  within the framework of the atomic simulation environment (ASE), \cite{larsen2017atomic}
  using the available libraries and APIs:
  MatGL (versions 1.1.2 and 1.1.3), \cite{matgl}
  Simple DFT-D3 (versions 1.0.0 and 1.1.0), \cite{dftd3} and
  DFT-D4 (version 3.6.0). \cite{dftd4}
From the MatGL library,
  we utilized the pre-trained model M3GNet-MP-2021.2.8-PES.
The default values of the DFT-D3(BJ) and 
                          DFT-D4 parameters for PBE were used. \cite{grimme2011effect,caldeweyher2020extension}

A sensible combination of the M3GNet potential with the DFT-D3(PBE) and DFT-D4(PBE) models is warranted 
  \textit{only if} the error in M3GNet energies, relative to PBE energies, is significantly smaller than the dispersion energy contributions.
A large-scale benchmark study\cite{chen2022universal} on M3GNet reports a mean absolute error (MAE) of 
  35 meV per atom in the M3GNet energies, relative to the corresponding PBE/PBE+$U$ energies.
This error is notably reduced to MAE~$= 12$~meV per atom 
  when considering the subset of M$^{\text{V}}$Q$^{\text{VI}}$X$^{\text{VII}}$ compounds studied here.
We also provide a plot of the M3GNet energies $E_{\text{M3GNet}}$
  (obtained from the companion website\cite{matterverse} of Ref.~\onlinecite{chen2022universal})
  versus the PBE energies $E_{\text{PBE}}$
  (retrieved from the Materials Project\cite{jain2013commentary} via the \texttt{MPRester} API client)
  as supplementary material,
  which reveals the linear correlation $E_{\text{M3GNet}}/N = 0.995 (E_{\text{PBE}}/N) - 0.014$~meV,
  with $N$ denoting the number of atoms in the compound's unit cell.
On the other hand,
  the dispersion energy contributions are in the range of $\sim 100$--$300$ meV per atom,
  as will be seen in Sec.~\ref{s:res},
  justifying the addition of D3 and D4 corrections to M3GNet.

We performed crystal structure optimizations that allowed both the unit cell vectors and ionic positions to relax,
  as well as fixed-volume optimizations to obtain EOSs.
The optimizations used experimental structures downloaded from the COD website \cite{gravzulis2009crystallography}
  as CIFs (crystallographic information files) \cite{hall1991crystallographic}
  for the initial configurations.
Using functions from ASE (version 3.23.0), \cite{larsen2017atomic} we refined the symmetry of the initial structures and 
  applied constraints to preserve space group symmetry during the optimizations.
In addition, the \texttt{FrechetCellFilter} class was used to apply
  the convergence criterion (set by a single parameter \texttt{fmax} in ASE) simultaneously to the atomic forces and unit cell stresses.
To ensure convergence, we used ASE optimizers
  based on the FIRE (fast inertial relaxation engine) \cite{bitzek2006structural} and 
               L-BFGS (limited-memory Broyden-Fletcher-Goldfarb-Shanno) \cite{nocedal1980updating} algorithms in succession.
We began with structure optimization using FIRE,
  setting $\texttt{fmax}=0.05$ and the initial time step $\texttt{dt}=1.0$, and allowing a maximum number of steps $\texttt{steps}=1000$.
We then switched to L-BFGS and refined the optimization with $\texttt{fmax}=0.02$ and $\texttt{steps}=100$.
Finally, we switched back to FIRE with $\texttt{fmax}=0.01$, $\texttt{dt}=0.5$, and $\texttt{steps}=2000$.
This strategy resulted in a fairly reasonable accuracy with the following values 
  for the maximum residual atomic force and stress components ($f_{\text{\tiny max}}$ and $\sigma_{\text{\tiny max}}$, respectively)
  in the optimized structures:
  $f_{\text{\tiny max}} \leq 9\ (16)$~meV/\AA\ for the optimized crystal structures used to obtain the EOS of BiTeI (BiTeBr);
  $f_{\text{\tiny max}} \leq 10$~meV/\AA\ and $\sigma_{\text{\tiny max}} \leq 0.11$~GPa for the rest of the optimized crystal structures,
  regardless of the potential used.

Following the crystal structure optimizations, 
  we calculated 
  the XRD patterns and RDF histograms from the CIF files of the optimized and experimental crystal structures
  using the functions
  \texttt{get\_rdf} from ASE \cite{larsen2017atomic} and 
  \texttt{XRDCalculator} from the Python Materials Genomics (pymatgen) library \cite{ong2013python} (version 2024.5.1), respectively.
The Earth mover's distance EMD$_{\text{XRD}}$ (EMD$_{\text{RDF}}$) between the XRD patterns (RDF histograms) of the optimized and corresponding experimental structures
  are computed using the function \texttt{wasserstein\_distance} from SciPy. \cite{virtanen2020scipy}
We prefer to use the EMD,
  which measures the minimum total ``work'' required to transform one distribution into another by treating intensity values as mass, 
  among other choices of dissimilarity measures, \cite{iwasaki2017comparison}
  because it does not require the peaks to occur in the same positions in the two distributions being compared and 
  accounts for intensities.
Higher values of EMD$_{\text{XRD}}$ (probing differences in interplanar distances)
             and EMD$_{\text{RDF}}$ (probing differences in interatomic distances) 
  indicate greater dissimilarity between the optimized and experimental structures,
  while lower values indicate greater similarity.
This overall approach for structure comparison is adopted
  because it is easily automated via scripting, 
             does not require significant human intervention, and 
             can be easily scaled to much larger datasets.

Among the CIF files obtained from the COD,
  we excluded the ones with partial site occupancies, and the ones where atoms were unrealistically close to each other,
  using PyCifRW\cite{hester2006validating} (version 4.4.6) parser and functions from pymatgen. \cite{ong2013python}
Furthermore, for the compounds SbSBr, SbSI, Sb$_4$O$_5$Cl$_2$, Sb$_4$O$_5$Br$_2$, BiOF, BiOCl, BiOBr, Bi$_4$O$_5$Br$_2$, BiOI, BiSCl, 
  among multiple CIF files belonging to the same compound with the same space group,
  only the one from the most recent experimental determination was used.
A list of the remaining 48 crystal structures, 
  along with clickable links to the information cards for entries on the COD website, \cite{gravzulis2009crystallography}
  is provided as supplementary material. 
Note that some crystal structures in the COD may require further experimental refinement,
  as distinguishing chalcogens from halogens can be challenging, 
  particularly in compounds with large unit cells or in the presence of disorder or non-stoichiometry, 
  where determining internal coordinates becomes increasingly complex.
Nevertheless, since our focus is on identifying general characteristics,
  we opted to use structures from the COD rather than manually collecting this data,
  which also aligns well with automation.

%
%
\section{\label{s:res}RESULTS AND DISCUSSION}

\begin{table}
  \caption{\label{t:bulkmod}
          The calculated and experimental values of
          the bulk modulus $K_0$ and its pressure derivatives 
                           $K_0^{\prime}$ and
                           $K_0^{\prime\prime}$
          of BiTeI and BiTeBr.
          }
  \begin{ruledtabular}
  \begin{tabular}{llD{.}{.}{1}D{.}{.}{1}D{.}{.}{1}l}
           & Method     &   \multicolumn{1}{c}{$K_0$} & \multicolumn{1}{c}{$K^\prime_0$} & \multicolumn{1}{c}{$K^{\prime\prime}_0$} & Reference \\ 
           &            &   \multicolumn{1}{c}{(GPa)} & \multicolumn{1}{c}{(GPa$^{-1}$)} & \multicolumn{1}{c}{(GPa$^{-1}$)}         &           \\ \hline
  BiTeI    &            &         &       &       & \\
           & M3GNet+D3  &   15.3  & 11.1  & -2.6  & \\
           & M3GNet+D4  &   17.6  & 10.6  & -2.2  & \\
           & DFT/PBEsol &   17.0  & 12.4  & -7.2  & \onlinecite{guler2016pressure}\\
           & Experiment &   20.5  & 7.6   &       & \onlinecite{xi2013signatures},~\onlinecite{guler2015crystal}\\
  BiTeBr   &                      &       &       & \\
           & M3GNet+D3  &   20.0  & 11.5  & -4.9  & \\
           & M3GNet+D4  &   22.1  & 10.9  & -4.0  & \\
           & DFT/PBEsol &   21    & 8.1   &       & \onlinecite{sans2016structural}\\
           & Experiment &   20    & 11    &       & \onlinecite{sans2016structural}\\
           &            &   22    & 7.5   &       & \onlinecite{ohmura2017pressure}\\
  \end{tabular}
  \end{ruledtabular}
\end{table}

Earlier DFT calculations for layered metal tellurohalides \cite{guler2015crystal} revealed that 
  the energy versus volume curve calculated using the PBE functional is anomalously flat near the equilibrium volume,
  particularly for volumes larger than the equilibrium value.
Additionally, the EOS calculated with PBE exhibits an excessively steep slope in the low-pressure region.
We find that these issues are also present in the M3GNet calculations 
  for BiTeBr [Figs.~\ref{f:ev}(a) and \ref{f:ev}(b)] as well as for BiTeI [Figs.~\ref{f:ev}(c) and \ref{f:ev}(d)].
This outcome is as expected since M3GNet was trained to reproduce PBE energies, forces, and stresses.
The comparison of M3GNet and PBE curves in Fig.~\ref{f:ev}(c) shows that the flatness of the M3GNet curve is inherited directly from PBE.
On the other hand, the M3GNet curve deviates from the PBE curve in the compression regime,
  leading to differences in the compressibility curves shown in Fig.~\ref{f:ev}(d).
This deviation is also expected 
  since M3GNet achieves a coefficient of determination \(R^2 = 0.757\) for bulk modulus predictions.\cite{chen2022universal}
For a subset of V-VI-VII compounds studied here,
  we calculate \(R^2 = 0.869\) and provide a plot of the Voigt-Reuss-Hill bulk moduli (M3GNet versus PBE) as supplementary material,
  demonstrating that M3GNet's accuracy does not deteriorate for V-VI-VII compounds.

In previous work, we found that adding a semiempirical dispersion (D2) force field \cite{grimme2006semiempirical} to PBE 
  substantially improved the prediction of compressibility. \cite{guler2015crystal}
Nevertheless, a more \textit{realistic} description of the experimental compressibility data for BiTeI as well as BiTeBr was achieved 
  using the PBEsol functional. \cite{guler2016pressure,sans2016structural}
Therefore, we will use results obtained with PBEsol, where available, 
  as the \textit{reference} for comparison with those obtained using the M3GNet+D3 and M3GNet+D4 potentials.
Another reason for this choice is that 
  the calculated and experimental EOSs usually refer to zero and room temperatures, respectively.
It is to be noted that 
  incorporating D3 corrections into PBE calculations yields optimized crystal structures that are highly comparable to those from PBEsol.
This is demonstrated by a plot of 
          the calculated\cite{rusinov2016pressure, monserrat2017temperature, brousseau2020temperature} lattice parameters 
  against the experimental\cite{shevelkov1995crystal, sans2016structural} lattice parameters for bismuth tellurohalides,
  provided as supplementary material,
  which shows that PBEsol results nevertheless align more closely with experimental values than PBE-D3.

As shown in Figs.~\ref{f:ev}(b) and \ref{f:ev}(d), the volume ($V$) vs. pressure ($P$) curves calculated 
  using the M3GNet+D3 and M3GNet+D4 potentials, rather than M3GNet alone, show much better agreement with the corresponding PBEsol curves.
This indicates that incorporating London dispersion significantly improves the prediction of compressibility.
The bulk modulus ($K_0$) and its pressure derivatives ($K_0^\prime$ and $K_0^{\prime\prime}$),
  obtained from Birch-Murnaghan (BM) fits, are presented in Table~\ref{t:bulkmod}.
Note that a third-order Birch-Murnaghan fit yields the values for $K_0$ and $K_0^\prime$, 
  while at least a fourth-order fit is necessary to determine $K_0^{\prime\prime}$.
Table~\ref{t:bulkmod} reveals that the overestimation of $K_0$ is balanced by the underestimation of $K_0^\prime$, and vice versa.
Additionally, discrepancies between experimental measurements for BiTeBr reported in Refs.~\onlinecite{sans2016structural,ohmura2017pressure}
  suggest that a reasonable error margin for $K_0$ is approximately 2 GPa.
Therefore, the results in Table~\ref{t:bulkmod} confirm that
  the accuracy of the dispersion-corrected M3GNet potentials is comparable to that of PBEsol.

\begin{figure}
  \includegraphics[width=0.45\textwidth]{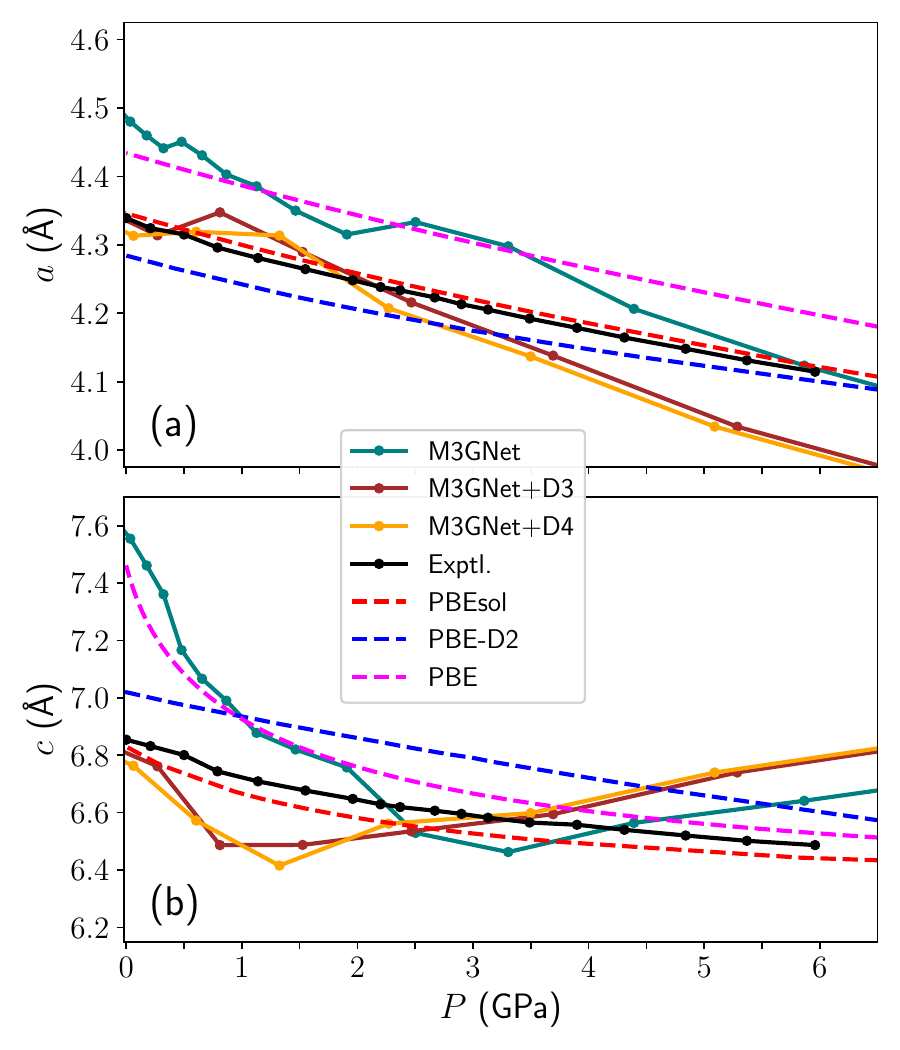}
  \caption{The variation of the lattice parameters $a$ (a) and $c$ (b) of BiTeI with the pressure $P$.
           The PBEsol, PBE-D2, and PBE curves are taken from Refs.~\onlinecite{guler2015crystal,guler2016pressure},
             while the experimental curves represent data from Ref.~\onlinecite{xi2013signatures}.}
  \label{f:acp}
\end{figure}

The graphs in Figs.~\ref{f:acp}(a) and \ref{f:acp}(b) show the variation of the lattice parameters $a$ and $c$ of BiTeI with pressure, respectively.
The steeper decay in the M3GNet $V(P)$ curve at low pressure is primarily due to the variation of $c$ with $P$.
This behavior is consistent with the PBE results.\cite{guler2015crystal}
However, it is important to note that the PBE-calculated $c(P)$ curve decreases monotonically,
  a trend not observed in the corresponding M3GNet curve.
This difference is likely due to the fact that the M3GNet potential was trained mostly on low-energy configurations.
Consequently, the improvement in crystal structure prediction through dispersion corrections becomes less effective for compressed structures.
This finding, in our opinion, unveils the importance of including at least moderately compressed (and expanded) and, 
  if feasible, high-pressure (and high-temperature) configurations in the training datasets of a MLIP.
For this reason, the remainder of the paper will focus on predicting equilibrium crystal structures at zero pressure.

\begin{figure*}
  \includegraphics[width=0.87\textwidth]{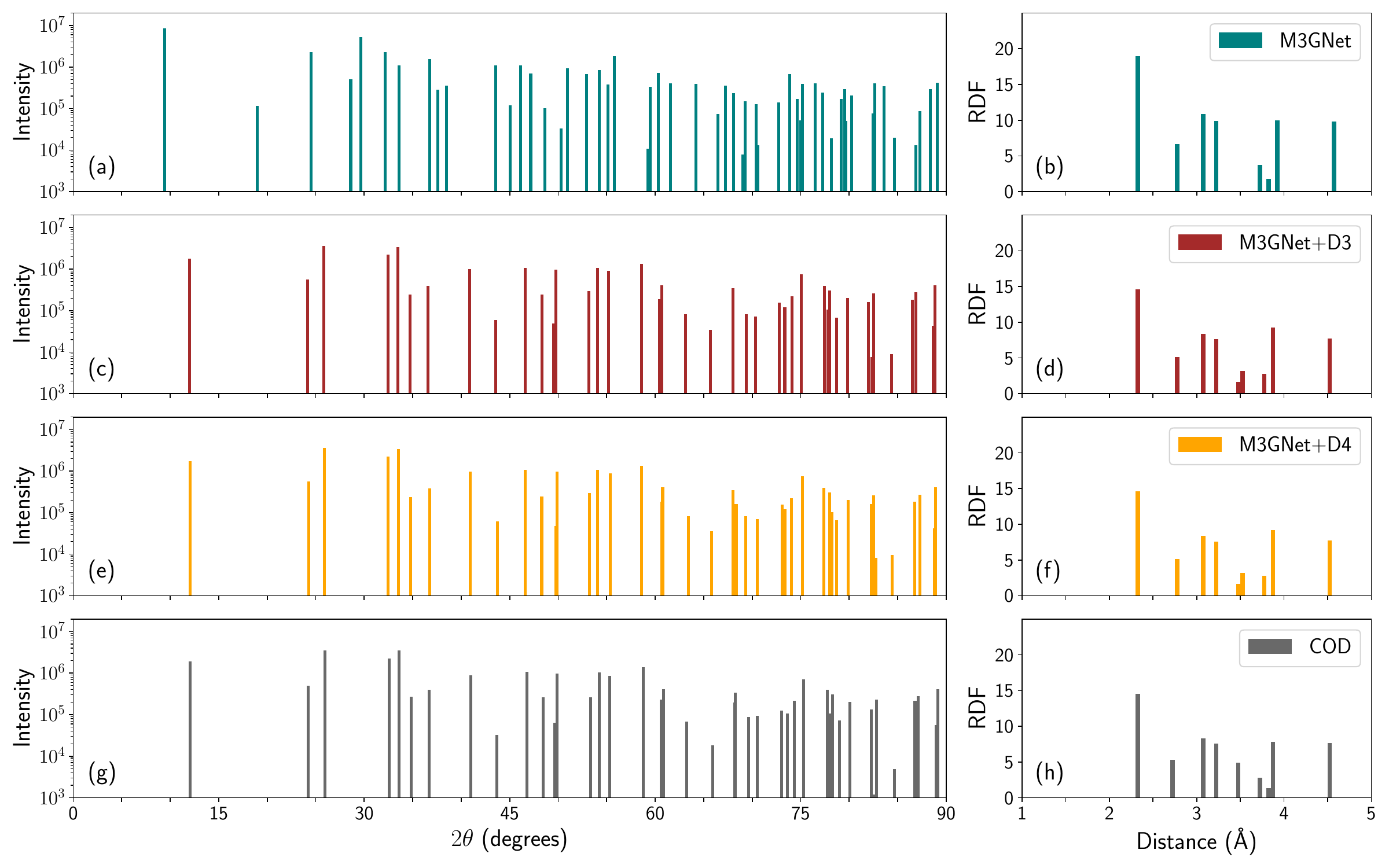}
  \includegraphics[width=0.87\textwidth]{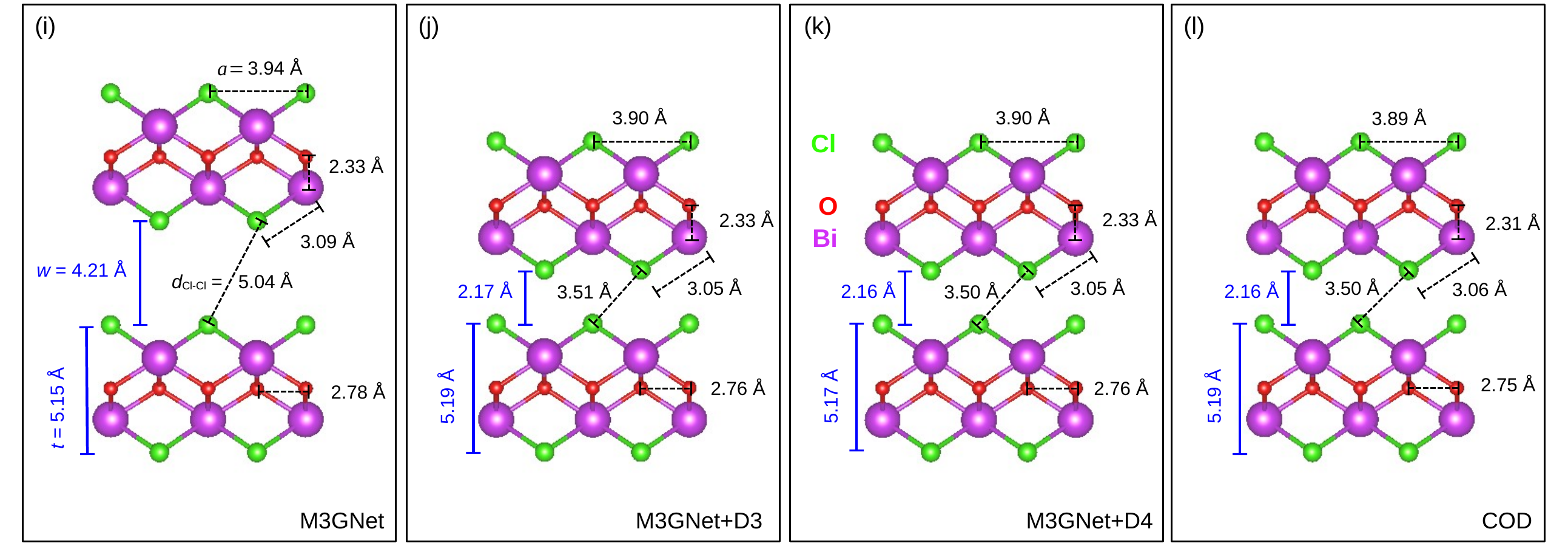}
  \includegraphics[width=0.87\textwidth]{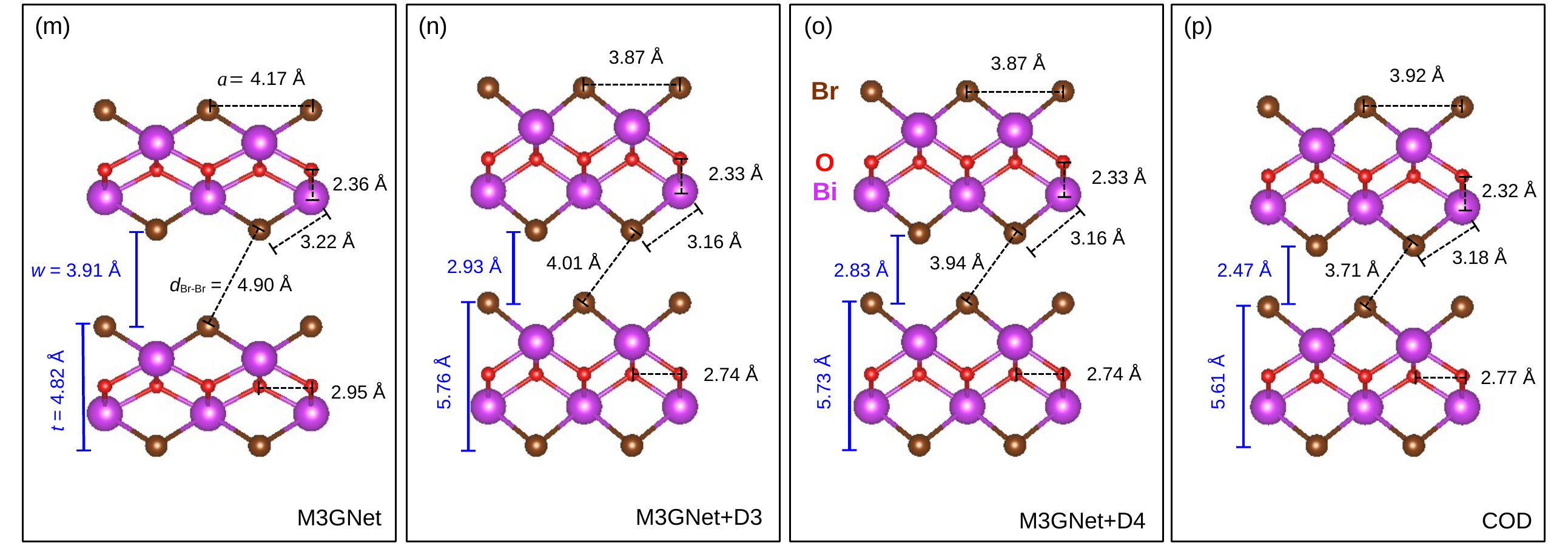}
  \caption{The XRD patterns calculated for BiOCl using the (a) M3GNet, (c) M3GNet+D3, (e) M3GNet+D4, and (g) COD crystal structures,
           displayed in (i), (j), (k) and (l), respectively, 
           with corresponding RDF histograms shown in (b), (d), (f), and (h).
           The optimized (M3GNet, M3GNet+D3, M3GNet+D4) and experimental (COD) crystal structures for BiOBr are drawn in (m), (n), (o), (p), respectively.
           The vdW gap and layer thickness are denoted as $w$ and $t$, respectively, with their sum equal to the lattice parameter $c$, i.e., $c = w + t$.
           Note that the lattice parameter $a$ and
           the interatomic distance between two chalcogen atoms 
           residing at the bottom of an upper layer and at the top of a lower layer ($d_{\text{Cl}-\text{Cl}}$ and $d_{\text{Br}-\text{Br}}$)
           are also indicated.
           }
  \label{f:xrdrdf}
\end{figure*}

\begin{figure}
  \includegraphics[width=0.45\textwidth]{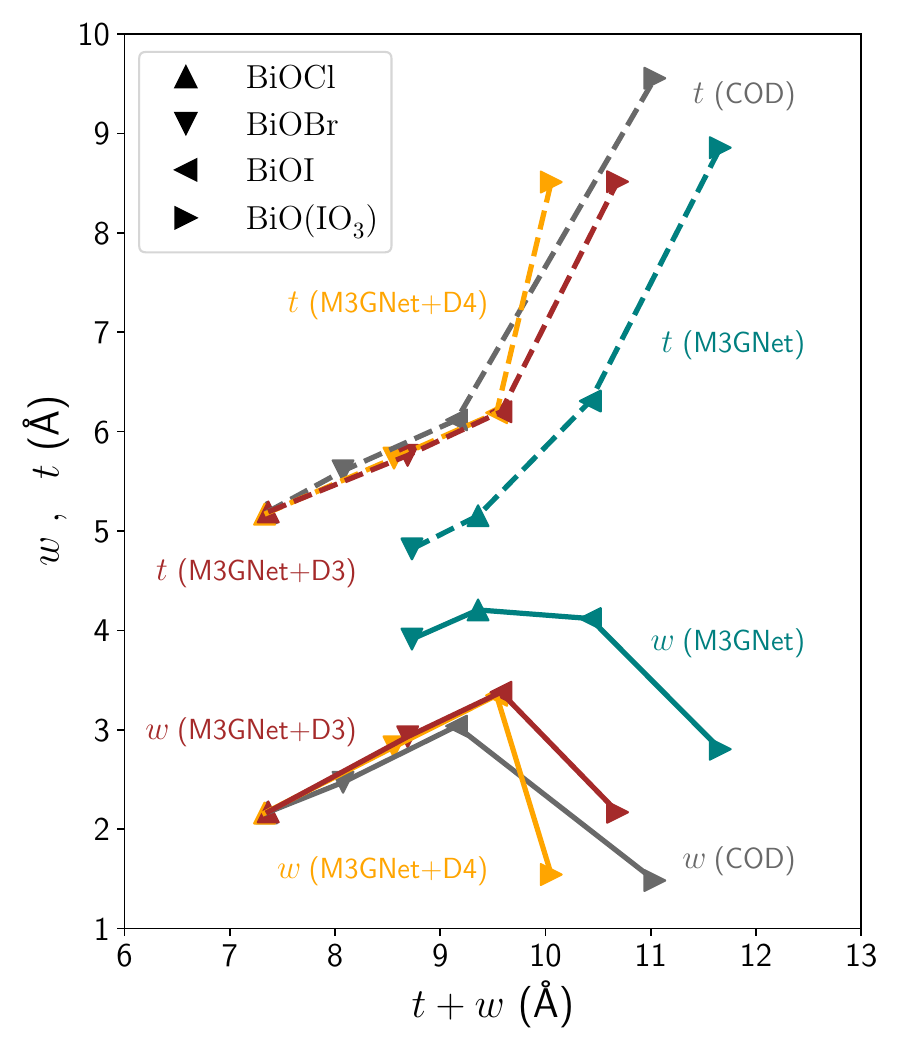}
  \caption{The plots showing the layer thickness ($t$) and the width of the vacuum region between layers ($w$)
             as a function of the lattice parameter perpendicular to the layer planes ($t + w$)
             for bismuth oxyhalides BiOX, where X = Cl, Br, and I, and BiO(IO$_3$).
           The lines are guides for the eye.}
  \label{f:vdW}
\end{figure}

We now focus on using the Earth mover's distances EMD$_{\text{XRD}}$ and EMD$_{\text{RDF}}$,
  computed from XRD patterns and RDF histograms, respectively, 
  to compare optimized crystal structures with experimental equilibrium structures from the COD.
The XRD patterns and RDF histograms for the 48 V-VI-VII compounds studied here are provided as supplementary material.
Larger (smaller) values of EMD$_{\text{XRD}}$ and EMD$_{\text{RDF}}$ indicate greater dissimilarity (similarity) 
  between the optimized and experimental crystal structures.
For BiOCl, the EMD$_{\text{XRD}}$ values are: 
  6.00 (M3GNet), 
  0.29 (M3GNet+D3), and 
  0.27 (M3GNet+D4).
Consistent with these values, 
  the XRD patterns in Figs.~\ref{f:xrdrdf}(c) and \ref{f:xrdrdf}(e) are noticeably more similar, 
  while that in Fig.~\ref{f:xrdrdf}(a) is less similar, 
  to the reference pattern in Fig.~\ref{f:xrdrdf}(g). 
Similarly, the EMD$_{\text{RDF}}$ values for this compound are:
  0.071 (M3GNet),
  0.012 (M3GNet+D3), and
  0.012 (M3GNet+D4).
Reflecting this trend,
  the RDF histograms in Figs.~\ref{f:xrdrdf}(d) and \ref{f:xrdrdf}(f) are again noticeably more similar,
  while that in Fig.~\ref{f:xrdrdf}(b) is less similar,
  to the reference histogram in Fig.~\ref{f:xrdrdf}(h).
This improvement with the inclusion of D3 and D4 corrections can be elucidated by 
  the visual inspection of the crystal structures in Figs.~\ref{f:xrdrdf}(i)-\ref{f:xrdrdf}(l).
While there is a slight improvement in predicting the lattice parameter $a$,
  the prediction of $c$ improves substantially, 
  driven by a more accurate estimation of the vacuum region width between layers $w$ (i.e., the vdW gap)
  and the interatomic distance between two Cl atoms residing at the bottom of an upper layer and at the top of a lower layer ($d_{\text{Cl}-\text{Cl}}$).
The relatively accurate estimation of the interplanar distance $w$ and interatomic distance $d_{\text{Cl}-\text{Cl}}$
  with the inclusion of dispersion corrections renders the respective peaks in the XRD patterns and RDF histograms 
  more consistent with the reference peaks, 
  thereby reducing the dissimilarity with the experimental crystal structure and 
  leading to smaller values of EMD$_{\text{XRD}}$ and EMD$_{\text{RDF}}$.

The inclusion of dispersion corrections leads to improved predictions of layer thickness $t$ (as well as $w$), 
  as illustrated for another bismuth oxyhalide in Figs.~\ref{f:xrdrdf}(m)-\ref{f:xrdrdf}(p).
The layer thickness $t$ for BiOBr is significantly underestimated with the M3GNet potential alone, 
  and this underestimation becomes a slight overestimation with the dispersion-corrected M3GNet potentials.
As shown for the bismuth oxyhalides BiOX (X = Cl, Br, and I) and BiO(IO$_3$) in Fig.~\ref{f:vdW}, 
  adding D3 and D4 corrections to the M3GNet potential results in systematic improvements in predicting 
  not only the vdW gap but also the layer thickness.

\begin{figure*}
  \includegraphics[width=0.95\textwidth]{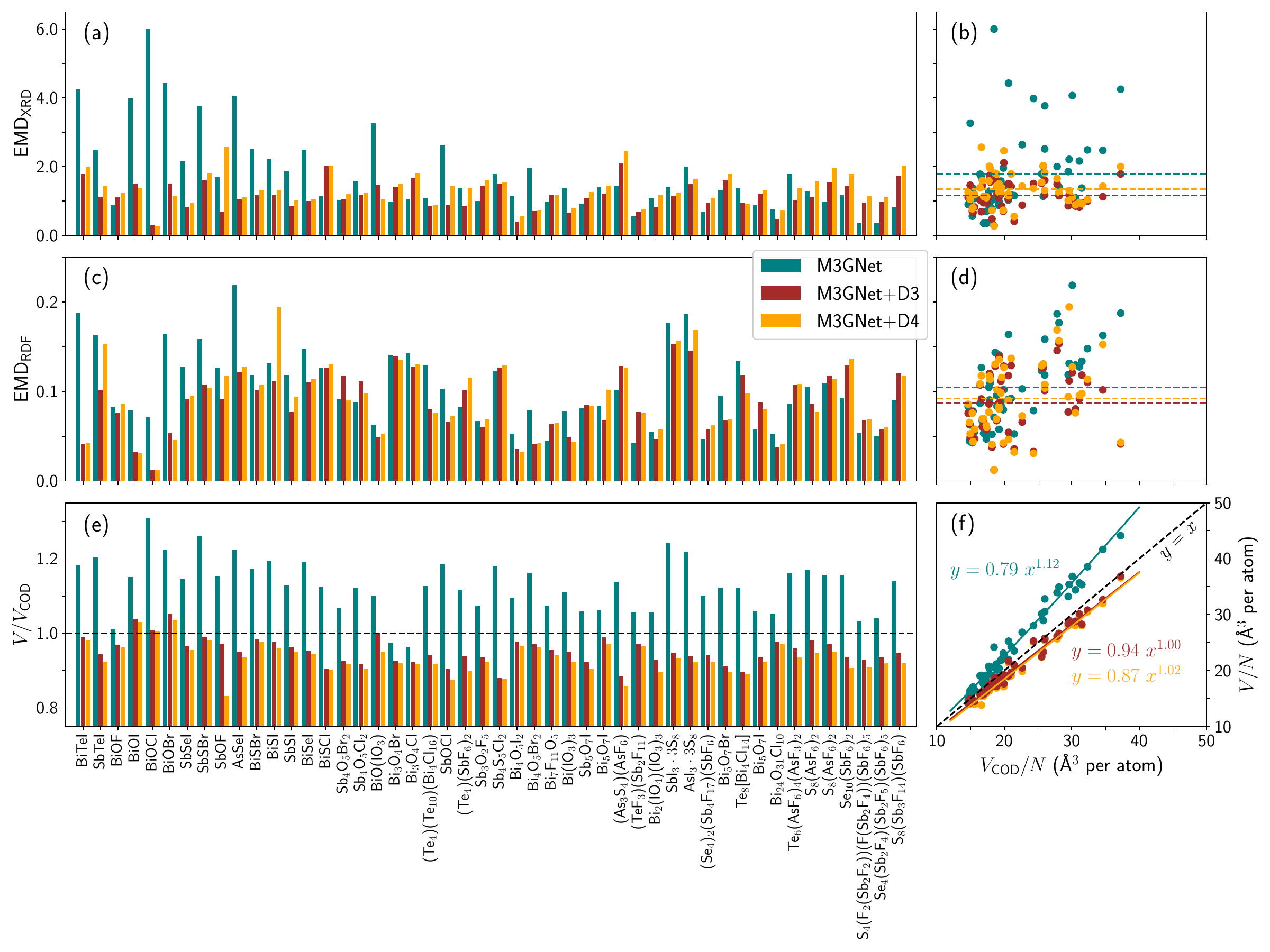}
  \caption{The bar plots showing the Earth mover's distances EMD$_{\text{XRD}}$ (a) and
                                                             EMD$_{\text{RDF}}$ (c),
                                               and the ratio $V/V_{\text{COD}}$ (e)
             for ternary M$^{\text{V}}$Q$^{\text{VI}}$X$^{\text{VII}}$ compounds (M = As, Sb, Bi; Q = O, S, Se, Te; X = F, Cl, Br, I),
             sorted by increasing number of atoms in the unit cell,
             where $V$ and $V_{\text{COD}}$ denote the optimized and experimental volumes, respectively.
           In the panels (b) and (d), EMD$_{\text{XRD}}$ and EMD$_{\text{RDF}}$ are plotted against $V_{\text{COD}}/N$, respectively,
             with the horizontal dashed lines indicating the mean values of EMD$_{\text{XRD}}$ and EMD$_{\text{RDF}}$.
           In the panel (f), $V/N$ is plotted against $V_{\text{COD}}/N$,
             with the solid lines representing the power-law regressions.
           }
  \label{f:ead}
\end{figure*}

Next, we compare the results of crystal structure optimizations using M3GNet, M3GNet+D3, and M3GNet+D4
  with the experimental equilibrium structures (COD) for the 48 aforementioned V-VI-VII compounds.
Figures~\ref{f:ead}(a) and \ref{f:ead}(c) show the bar plots of EMD$_{\text{XRD}}$ and EMD$_{\text{RDF}}$
  for these compounds.
The most significant improvement from incorporating London dispersion is observed for BiOCl discussed above.
The reduction in EMD$_{\text{XRD}}$ resulting from the addition of D3 and D4 corrections to M3GNet is more pronounced 
  compared to the reduction in EMD$_{\text{RDF}}$.
This indicates that the D3 and D4 corrections have a greater impact on interactions with longer ranges, rather than shorter ranges, 
  which aligns with the physical insights behind the construction of DFT-D3 and DFT-D4 models.

The mean EMD$_{\text{XRD}}$ [EMD$_{\text{RDF}}$] values are 
  1.8 [0.104], 
  1.2 [0.087], and
  1.3 [0.092] for M3GNet, M3GNet+D3, and M3GNet+D4, respectively,
  as indicated by the dashed lines in Fig.~\ref{f:ead}(b) [\ref{f:ead}(d)].
Lower values of EMD$_{\text{XRD}}$ and EMD$_{\text{RDF}}$ indicate greater similarity between the optimized and experimental crystal structures.
However, it should be noted that for some compounds,
  especially those with relatively large unit cells containing a greater number of atoms, 
  both EMD$_{\text{XRD}}$ and EMD$_{\text{RDF}}$ increase when the D3 and D4 corrections are applied.
This means that the improvement with combined potentials is not universal.
On the other hand, it should also be noted that experimentally determining the internal coordinates becomes more difficult as $N$ increases,
  which leads to uncertainties in the experimental structures.
Since the unit cell volume is usually measured with higher accuracy,
  we compare the optimized and experimental volumes ($V$ and $V_{\text{COD}}$, respectively) to each other in Figs.~\ref{f:ead}(e) and \ref{f:ead}(f). 
It is seen that while M3GNet substantially overestimates the unit cell volume, 
  the inclusion of D3 and D4 corrections in M3GNet results in a slight underestimation.
This behavior is quantified by the power-law regressions in Fig.~\ref{f:ead}(f) for $V/N$ versus $V_{\text{COD}}/N$,
  i.e., $V/N=A(V_{\text{COD}}/N)^\alpha$.
The coefficients of regression are given in Table~\ref{t:powreg}.
Typically, either the prefactor $A$, the exponent $\alpha$, or both in such regressions should be slightly less than unity,
  because the optimized volumes refer to zero temperature, while the experimental volumes correspond to finite temperatures.
This expectation is best satisfied by the optimized volumes obtained with M3GNet+D3,
  as seen in Table~\ref{t:powreg},
  even though the regression lines of M3GNet+D3 and M3GNet+D4 are almost indistinguishable in Fig.~\ref{f:ead}(f).
It is therefore clear that, on average, the dispersion-corrected M3GNet potentials provide superior volume predictions compared to M3GNet alone.
Among the two dispersion-corrected potentials, M3GNet+D3 appears to be superior to M3GNet+D4.

The prefactor $A$ takes a significantly smaller value for M3GNet+D4 compared to M3GNet+D3.
This suggests that the van der Waals binding energy is overestimated with M3GNet+D4, 
  given that the volume estimates are more realistic with M3GNet+D3 and 
  overbinding generally leads to smaller equilibrium volumes.
In order to quantify this observation,
  we explore the correlation between the energy differences 
  $\Delta_4=(E_{\text{min}}^{\text{M3GNet+D4}}-E_{\text{min}}^{\text{M3GNet}})/N$ and 
  $\Delta_3=(E_{\text{min}}^{\text{M3GNet+D3}}-E_{\text{min}}^{\text{M3GNet}})/N$,
  where $E_{\text{min}}^{\text{P}}$ denotes the minimum energy at the equilibrium volume calculated with potential $\text{P}$.
The plot of $\Delta_4$ versus $\Delta_3$ is provided as supplementary material. 
Both $\Delta_3$ and $\Delta_4$ take values in the range of $\sim 100$--$300$ meV per atom 
  for the M$^{\text{V}}$Q$^{\text{VI}}$X$^{\text{VII}}$ compounds,
  as mentioned in Sec.~\ref{s:met}.
We find
  $\Delta_4=1.13\ \Delta_3 - 0.02$ eV,
  indicating that the van der Waals binding energy is overestimated by about 13\% with M3GNet+D4 relative to M3GNet+D3.
This overestimation explains why M3GNet+D4 leads to a more pronounced underestimation of the unit cell volume compared to M3GNet+D3.

\begin{table}
  \caption{\label{t:powreg}The prefactor $A$ and exponent $\alpha$ values from power-law regression $V/N=A(V_{\text{COD}}/N)^\alpha$.}
  \begin{ruledtabular}
  \begin{tabular}{lD{.}{.}{2}D{.}{.}{2}}
     Potential &   \multicolumn{1}{c}{$A$} & \multicolumn{1}{c}{$\alpha$}  \\ \hline
     M3GNet    &   0.79 & 1.12 \\
     M3GNet+D3 &   0.94 & 1.00 \\
     M3GNet+D4 &   0.87 & 1.02 
  \end{tabular}
  \end{ruledtabular}
\end{table}

%
%
\section{\label{s:con}Conclusion}

In this work, we explored the idea of combining a universal graph deep learning interatomic potential (M3GNet),
  trained on datasets from density-functional calculations using a semi-local exchange-correlation functional,
  with two generally applicable London dispersion correction models (DFT-D3 and DFT-D4).
Our objective was to assess whether dispersion-corrected M3GNet potentials can provide an adequate physical description of pnictogen chalcohalide compounds
  without fine-tuning the training or refitting the potential parameters.

We first derived the equations of state for layered pnictogen chalcohalides BiTeBr and BiTeI, and found that
  the accuracy of the M3GNet+D3 and M3GNet+D4 potentials is comparable to that of a highly reliable density functional.
We then optimized and characterized the crystal structures of a broader set of V-VI-VII compounds.
To quantify the dissimilarity between the optimized and corresponding experimental structures, 
  we utilized two Earth Mover's distances computed from X-ray diffraction patterns and radial distribution function histograms.
Our results indicate that
  dispersion-corrected potentials generally provide a more realistic description of the crystal structures of V-VI-VII compounds
  due to the inclusion of van der Waals attractions,
  compared to using the graph deep learning potential alone.

The improvement resulting from incorporating London dispersion corrections becomes less effective for compressed structures,
  which could hinder the broader applicability of the M3GNet+D3 and M3GNet+D4 potentials.
We believe this limitation can be addressed 
  by including compressed and expanded (high-pressure and high-temperature) configurations
  in the dataset used to train the graph deep learning potential.
Additionally, for some compounds, 
  especially those with relatively large unit cells containing a greater number of atoms,
  the addition of dispersion corrections does not improve the description of crystal structures, 
  indicating that the improvement with combined potentials is \textit{not universal}.
On the other hand, while the unit cell volume
  is substantially overestimated with the graph deep learning potential, 
  this error turns into a slight underestimation when dispersion corrections are applied.
Notably, incorporating London dispersion corrections
  leads to systematic improvements in predicting not only the van der Waals gap but also layer thickness in layered V-VI-VII compounds.
While it remains a question for future studies 
  whether the combined potentials studied here can be made universal and broadly applicable through fine-tuning and refitting,
  we hope that, in their current simple and ready-to-use form, 
  they will enable cost-effective and realistic atomistic simulations of an important class of materials (layered pnictogen chalcohalides), 
  systems derived from them, and composites made of them, thereby facilitating further theoretical studies on these systems.

%
%
\section*{Supplementary Material}

Supplementary material is available online as a single PDF containing
  a plot of M3GNet versus PBE energies,
  a list of V-VI-VII compounds studied in the paper, with clickable links to their information cards on the COD website,
  a plot of the Voigt-Reuss-Hill bulk moduli (M3GNet versus PBE),
  the X-ray diffraction patterns and radial distribution function histograms calculated for the optimized and experimental crystal structures, 
  a plot of the calculated lattice parameters against the experimental lattice parameters for bismuth tellurohalides, and 
  a plot of $\Delta_4$ versus $\Delta_3$.

%
%
\begin{acknowledgments}
The computations reported here were partially performed at the High Performance and Grid Computing Center (TRUBA Resources) of TUBITAK ULAKBIM.
The atomistic structures in Fig.~\ref{f:xrdrdf} were generated using the visualization software VESTA 3.\cite{momma2011vesta}
\end{acknowledgments}

%
%
\section*{AUTHOR DECLARATIONS}
\subsection*{Conflict of Interest}
The authors have no conflicts to disclose.

\subsection*{Author Contributions}
\noindent
\textbf{\c{C}etin K{\i}l{\i}\c{c}:} Conceptualization (lead);
                                    Formal analysis (lead);
                                    Investigation (lead); 
                                    Methodology (lead); 
                                    Visualization (equal);
                                    Writing -- original draft (lead); 
                                    Writing -- review \& editing (lead).
\textbf{S\"{u}meyra G\"{u}ler-K{\i}l{\i}\c{c}:} Formal analysis (supporting);
                                                Investigation (supporting); 
                                                Visualization (equal);
                                                Writing -- review \& editing (supporting).

%
%
\section*{DATA AVAILABILITY}
The data that supports the findings of this study are available
  within the article and its supplementary material.


\nocite{*}
\bibliography{refs}

\end{document}